\documentclass[prb,superscriptaddress,floatfix,tightenlines,10pt,twocolumn]{revtex4}
\usepackage{amssymb}
\usepackage{amsmath}
\usepackage{graphicx}

\begin{document}

\title{Nuclear magnetic resonance\ line shapes of Wigner crystals in $^{13}$%
C-enriched graphene}
\author{R. C\^{o}t\'{e}}
\affiliation{D\'{e}partement de physique and Institut quantique, Universit\'{e} de
Sherbrooke, Sherbrooke, Qu\'{e}bec, J1K 2R1, Canada}
\author{Jean-Michel Parent}
\affiliation{D\'{e}partement de physique and Institut quantique, Universit\'{e} de
Sherbrooke, Sherbrooke, Qu\'{e}bec, J1K 2R1, Canada}

\begin{abstract}
Assuming that the nuclear magnetic resonance (NMR) signal from a $^{13}$C
isotope enriched layer of graphene can be made sufficiently intense to be
measured, we compute the NMR\ lineshape of the different crystals ground
states that are expected to occur in graphene in a strong magnetic field. We
first show that in nonuniform states, there is, in addition to the frequency
shift due to the spin hyperfine interaction, a second contribution of equal
importance from the coupling between the orbital motion of the electrons and
the nuclei. We then show that, if the linewidth of the bare signal can be
made sufficiently small, the Wigner and bubble crystals have line shapes
that differ qualitatively from that of the uniform state at the same density
while crystal states that have spin or valley pseudospin textures do not.
Finally, we find that a relatively small value of the bare linewidth is
sufficient to wash out the distinctive signature of the crystal states in
the NMR line shape.
\end{abstract}

\date{\today }
\maketitle

\section{INTRODUCTION}

In a perpendicular magnetic field $\mathbf{B}=B_{0}\widehat{\mathbf{z}}$,
the eigenstates of the two-dimensional electron gas (2DEG) are quantized
into Landau levels with energy $E_{n}=(n+1/2)\hslash \omega _{c},$ where $%
n=0,1,2,...$ is the Landau level index and $\omega _{c}=eB_{0}/m^{\ast }$ is
the cyclotron frequency with $m^{\ast }$ the effective electronic mass. At
integer fillings of these levels, the 2DEG exhibits the quantum Hall effect%
\cite{ReviewQHE}. However, for non-integer fillings and in certain ranges of
filling factor $\nu =N_{e}/N_{\varphi }$ in each Landau level ($N_{e}$ is
the number of electrons and $N_{\varphi }=S/2\pi \ell ^{2}$ is the Landau
level degeneracy with $S$ the area of the 2DEG and $\ell =\sqrt{\hslash
/eB_{0}}$ the magnetic length), the electrons can form a Wigner crystal\cite%
{Wigner}\textit{\ }i.e., a crystal of electrons. In Landau levels $n>1$, the
Coulomb interaction can also favor the formation of bubble crystals (i.e.,
Wigner crystals with more than one electron per site\cite%
{Revuebulles,Wignerrevue}) or, near half-integer filling, a the formation of
a stripe phase\cite{Stripe}). In GaAs quantum wells, these crystal states
have been detected by a series of transport and microwave absorption
experiments\cite{Microwave}. More recently, resistively-detected nuclear
magnetic resonance (RD-NMR) has been used to study the Wigner and bubble
crystals in Landau levels $n=0,1$ in these quantum wells\cite{Tiemann}. This
technique uses the strong hyperfine contact interaction between the electron
and the nuclear spins in GaAs quantum wells to obtain a signal. In crystal
states, the spatial modulation of the electronic density leads to a
modulation of the hyperfine coupling and thus to a modulation of the nuclear
frequency shift (the Knight shift). In consequence, the effective electronic
Zeeman energy, which is what is ultimately detected in RD-NMR, also varies
spatially. The resulting NMR\ signal is a reflection of the electronic
density distribution in the crystal\cite{Tiemann}. From the spectral line
shape of the NMR signal, it is thus possible to distinguish between
different types of crystal states\cite{Tiemann,CoteNMR}.

Magnetically induced Wigner crystal were also predicted to occur in graphene%
\cite{CoteSkyrmeGraphene,GoerbigCW,Joglekar} and in multilayer graphene\cite%
{Yoshioka}. In graphene, the Landau level energies are given by $E_{n}=%
\mathrm{sgn}\left( n\right) \sqrt{2\hslash v_{F}^{2}e\left\vert n\right\vert
B_{0}},$ where the Landau level index $n=0,\pm 1,\pm 2,...$ and the Fermi
velocity $v_{F}=3c\gamma _{0}/2\hslash ,$ with $c=1.42$ \AA\ the separation
and $\gamma _{0}=3.12$ eV the hopping energy between adjacent carbon atoms.
Since the Zeeman coupling is very small, Landau levels are usually
considered as fourfold degenerate when counting spin and valley degrees of
freedom. The additional valley degeneracy in graphene increases the
diversity of crystal phases. For example, crystals with a valley pseudospin
texture become possible ground states. In this paper, we consider, from a
theoretical point of view, the possibility to detect these crystals using
their signal in bulk NMR. We are motivated by the chiral nature of the
conduction electrons in graphene which, as was shown recently\cite{Dora},
leads to a NMR behavior that is different from that of a normal metal. For
example, we show that, when nonuniform states are considered, the
interaction between the nuclear spin of a carbon $^{13}$C atom and the
orbital motion of the conduction electrons is nonzero. It is of the same
order as the hyperfine coupling between a nuclear spin and the spin of the
conduction electrons. In graphene, both couplings cause a NMR\ frequency
shift. They must be summed to compute the total signal. We study the line
shape of the crystal phases predicted theoretically and compare them with
the NMR signal obtained from a uniform state with the same average density
of electrons. Among the predicted ground states, we find that Wigner and
bubble crystals (i.e., states with no valley and/or spin texture) have NMR\
line shapes that differ qualitatively from those of the corresponding
uniform states in part because the orbital shift is zero in the uniform
states. Thus, in principle, bulk NMR could be used to identify the presence
of the crystal states in the phase diagram of the chiral 2DEG in graphene.

Nuclear magnetic resonance\ cannot be used for ordinary graphene since the $%
^{12}$C atoms have no nuclear spin and the percentage of $^{13}$C atoms
(which have a nuclear spin) is very small (of the order of $1\%$). It is
possible, however, to increase the fraction of $^{13}$C atoms in graphene
sheets and even to fabricate pure $^{13}$C graphene. Even then, the NMR\
signal may be too small to allow detection by bulk NMR\ since the number of
nuclei is small in a two-dimensional system unless a prohibitively large
sheet is considered. This problem was addressed previously in Ref. %
\onlinecite{Dora} where it was concluded that bulk NMR detection in $^{13}$%
C-enriched graphene sheets should soon be within experimental reach. We will
assume that such is the case and concern ourselves solely with the width and
shape of the spectral line. We remark that RD-NMR would be another way to
solve the intensity problem since it is more adapted to system with a small
density of nuclear spins\cite{Gervais}.

Although there is a number of theoretical calculations of NMR parameters in
carbon nanotubes and graphene\cite{Fabio}, there are at present no available
NMR\ experimental results for a single graphene layer so that we have to
rely on estimates for the NMR parameters. For the hyperfine contact and
dipolar interactions, we use the values given in Ref. \onlinecite{Dora}. For
the bare NMR\ linewidth which is also unknown, our calculation allows us, by
comparing the signal from the crystal and uniform state, to estimate the
maximal value above which it is no longer possible to distinguish a given
crystal state from the corresponding uniform state.

Our paper is organized as follow. We explain in Sec. II how the NMR line
shape is calculated when crystal states are considered. In Sec. III, we
derive the Hamiltonian for the coupling between the nuclear and electron
spin and summarize our calculation for the coupling between the nuclear spin
and the orbital motion of the electrons. In Sec. IV, we review how the
calculation of the Fourier components of the electronic and spin densities
of the crystals is done. The line shapes of the predicted crystal states in
graphene are calculated in Sec. V. We conclude in Sec. VI.

\section{NMR LINE SHAPE FOR CRYSTAL STATES}

Graphene is a single sheet of carbon atoms arranged in a honeycomb lattice
which can be described as a triangular Bravais lattice with a basis of two
carbon atoms denoted $A$ and $B$. The triangular lattice constant $a_{0}=%
\sqrt{3}c,$ where $c=1.42$ \AA\ is the separation between to adjacent carbon
atoms. The Brillouin zone of the reciprocal lattice has $6$ valleys, but
only two of them are inequivalent. We use $\xi =\pm 1$ for the valley index
and choose the two inequivalent valleys to be at wave vectors $\mathbf{K}%
_{\xi }=\frac{2\pi }{a_{0}}\xi \left( \frac{2}{3},0\right) $. In the
sublattice basis $\left( A,B\right) ,$ and for wave vectors $\mathbf{k}$
near $\mathbf{K}_{\xi }$ (i.e., in the \textit{continuum} approximation),
the noninteracting electronic Hamiltonian is given by%
\begin{equation}
H=\xi \hslash \nu _{F}(\tau _{x}k_{x}-\xi \tau _{y}k_{y}),  \label{rene3}
\end{equation}%
where $v_{F}=3c\gamma _{0}/2\hslash =9.\,\allowbreak 06\times 10^{5}$ m$%
\cdot $s$^{-1}$ is the Fermi velocity with $\gamma _{0}=2.8$ eV the
nearest-neighbor hopping energy. In Eq. (\ref{rene3}), $\tau _{x},\tau _{y}$
are Pauli matrices in sublattice space (we later use $\mathbf{\sigma }$ for
the Pauli matrices in spin space). It follows from Eq. (\ref{rene3}) that in
the continuum approximation, electrons are massless and chiral\cite%
{GrapheneReview}.

In a perpendicular magnetic field $\mathbf{B}=B_{0}\widehat{\mathbf{z}},$
the kinetic energy is quantized into Landau levels with energy $E_{n}=%
\mathrm{sgn}\left( n\right) \sqrt{2\hslash v_{F}^{2}e\left\vert n\right\vert
B_{0}}$ where the Landau level index $n=0,\pm 1,\pm 2,...$ and $\mathrm{sgn}$
is the signum function. Each Landau level $n$ contains four sublevels with
quantum numbers $\left( \xi ,\sigma \right) ,$ where $\sigma =\pm 1$ is the
spin index. In the absence of Coulomb interaction, these four levels are
almost degenerate owing to the smallness of the Zeeman coupling which is
given by%
\begin{equation}
g\mu _{B}B_{0}=0.1\,\allowbreak 16B_{0}\text{ meV,}  \label{zeeman}
\end{equation}%
with $B_{0}$ in Tesla and a g-factor given by $g=2.$ For Landau level $n=1$, 
$E_{1}=32.88\sqrt{B_{0}}$ meV while the Coulomb energy $e^{2}/4\pi
\varepsilon _{r}\varepsilon _{0}\ell =22.91\sqrt{B}$ meV for $\varepsilon
_{r}=2.45$ appropriate for graphene on a SiO$_{2}$ substrate\cite{Jang}.

We denote the filling factor of level $n$ by $\nu _{n}=2\pi n_{e}\ell
^{2}\in \left[ 0,4\right] ,$ where $n_{e}$ is the averaged electronic
density \textit{in this level. }The average spin density (in the direction
of the applied magnetic field) is given by%
\begin{equation}
\left\langle S_{z}\left( \mathbf{r}\right) \right\rangle =\frac{1}{2}\hslash %
\left[ \left\langle n_{+}\left( \mathbf{r}\right) \right\rangle
-\left\langle n_{-}\left( \mathbf{r}\right) \right\rangle \right] ,
\end{equation}%
where $\left\langle n_{\sigma }\left( \mathbf{r}\right) \right\rangle $ is
the areal density of each spin species and $\left\langle \cdots
\right\rangle $ denotes a ground state average (we set $T=0~$K in all our
calculations).

Following previous works\cite{CoteNMR,Tiemann}, we write $g\left( f\right) $
for the intrinsic line shape of the $^{13}$C nucleus and assume that it has
the Gaussian form 
\begin{equation}
g\left( f\right) =A_{0}e^{-\left( f-f_{0}\right) ^{2}/\Gamma ^{2}},
\label{gamma}
\end{equation}%
where $\Gamma $ is the intrinsic linewidth, $f_{0}$ the bare resonance
frequency and $A_{0}$ the amplitude. Since we are not interested in the
absolute amplitude of the signal, we set $A_{0}=1$ in our calculations. The $%
^{13}$C isotope has spin $I=1/2$ and its nuclear magnetic moment is given by 
$\mathbf{\mu }_{n}=\hslash \gamma _{n}\mathbf{I}_{n},$ where the
gyromagnetic factor $\gamma _{n}/(2\pi )=10.705$ MHz/T.\ In an external
magnetic field $B_{0},$ the bare\ nuclear precession frequency is $%
f_{0}=\left( \gamma _{n}/2\pi \right) B_{0}.$ This frequency is shifted by
the interaction $h_{D}$ between the $^{13}$C nuclei and the electron gas.
The difference between the measured and bare frequencies is $\Delta f\left( 
\mathbf{r}\right) $. We compute the interaction $h_{D}$ in the next section
and show that the resulting frequency shift is proportional to the
electronic density for the coupling with the orbital motion of the electrons
and to the spin density for the Knight shift. The line shape $I\left(
f\right) $ of the crystal state is obtained by summing the contribution from
all nuclei in the graphene sheet. Since we work in the continuum
approximation, we assume that these nuclei form a continuous background and
so the intensity per carbon atom is given by%
\begin{equation}
I\left( f\right) =\frac{2}{N_{c}A_{c}}\int d\mathbf{r}g\left( f-f_{0}-\Delta
f\left( \mathbf{r}\right) \right) ,
\end{equation}%
where $A_{c}$ is the area of a primitive unit cell of the graphene lattice
(which contains two carbon atoms), $N_{c}$ is the total number of carbon
atoms and the integral is over the graphene sheet. The spatial variations of
the electronic density in a non uniform state leads to an inhomogeneous
broadening of the intrinsic line shape. In this paper, we concentrate on
Wigner crystal phases so that $\Delta f\left( \mathbf{r}\right) $ is a
periodic function. It is then only necessary to integrate over a unit cell
area ($A_{WC}$) of the Wigner crystal i.e.,%
\begin{equation}
I\left( f\right) =\frac{1}{A_{WC}}\int_{WC}d\mathbf{r}g\left( f-f_{0}-\Delta
f\left( \mathbf{r}\right) \right) .  \label{rene7}
\end{equation}%
The frequency shift $\Delta f\left( \mathbf{r}\right) =\Delta f_{orb}\left( 
\mathbf{r}\right) +\Delta f_{spin}\left( \mathbf{r}\right) $ has two
contributions, the first one, $\Delta f_{orb}\left( \mathbf{r}\right) ,$
comes from the orbital motion of the electrons while the second one, $\Delta
f_{spin}\left( \mathbf{r}\right) ,$ comes from the interaction between the
electronic spin density and the spin of the nuclei.

\section{ELECTRON-NUCLEUS INTERACTIONS}

The interaction Hamiltonian $h_{D}$ is the sum of the two contributions%
\begin{equation}
h_{D}=h_{orb}+h_{spin},
\end{equation}%
where $h_{orb}$ is the coupling between a nucleus and the orbital motion of
an electron and $h_{spin}$ is the coupling between a nucleus and the
electronic spin density around it. The hyperfine coupling $h_{spin}$ is
itself the sum of a dipolar and a Fermi hyperfine contact interactions.
Although the $\pi -$bands conduction electrons in graphene occupy $p_{z}$
orbitals, the contact interaction is nevertheless not zero because of
contributions from the valence $\sigma -$band states which are due to
exchange-polarization effects\cite{Yazyev}.

Due to the massless character of the electron in graphene, the standard
textbook form\cite{Slichter} of the hyperfine interaction $h_{orb}$ needs to
be reformulated. As shown in Ref. \onlinecite{Dora}, the coupling between
the magnetic moment $\hslash \gamma _{n}I_{z}$ of a nuclei at $\mathbf{R}=0$
and the magnetic field $\mathbf{B}_{orb}\left( 0\right) $ due to the orbital
motion of one electron is given in graphene by 
\begin{equation}
h_{orb}\left( \mathbf{r}\right) =-\frac{\mu _{0}}{4\pi }\hslash \gamma
_{n}I_{z}\left( \frac{\mathbf{r}\times \mathbf{j}_{\xi ,\sigma }}{r^{3}}%
\right) _{z},  \label{couplingh}
\end{equation}%
where 
\begin{equation}
\mathbf{j}_{\xi ,\sigma }=-\xi e\nu _{F}\left[ \tau _{x}\widehat{\mathbf{x}}%
-\xi \tau _{y}\widehat{\mathbf{y}}\right]  \label{currentg}
\end{equation}%
is the electric current of an electron with valley and spin indices $\xi
,\sigma .$

The Hamiltonian $h_{spin}$ for an electron at $\mathbf{r}$ and an ion at $%
\mathbf{R}=0$ is unchanged with respect to its standard form and is given by%
\cite{Slichter}%
\begin{eqnarray}
h_{spin,\xi }\left( \mathbf{r}\right) &=&-\mathbf{\mu }_{e}\cdot \mathbf{B}%
_{n}  \label{rene2} \\
&=&\frac{\mu _{0}}{4\pi }g\mu _{B}\hslash \gamma _{n}\mathbf{I}  \notag \\
&&\cdot \left[ \frac{8\pi }{3}\mathbf{S}_{\xi }\delta \left( \mathbf{r}%
\right) -\frac{\mathbf{S}_{\xi }r^{2}-3\left( \mathbf{S}_{\xi }\cdot \mathbf{%
r}\right) \mathbf{r}}{r^{5}}\right] ,  \notag
\end{eqnarray}%
where (and in this expression only) $\mathbf{r}$ is a three-dimensional
vector and $\delta \left( \mathbf{r}\right) $ the three-dimensional Dirac
delta function. Moreover $\mathbf{B}_{n}$ is the magnetic field of the
nucleus of spin\ $\mathbf{I}$, $\mathbf{S}_{\xi }$ is the spin of an
electron in valley $\xi $ and $\mu _{0}$ is the vacuum permeability. The
Bohr magneton $\mu _{B}=e\hslash /2m_{e},$ where $m_{e}$ is the \textit{bare}%
\cite{Yafet} electronic mass\cite{Slichter}. The electron magnetic moment is 
$\mathbf{\mu }_{e}=-g\mu _{B}\mathbf{S}$ where $S=1/2$ is the electronic
spin ($\mathbf{I}$ and $\mathbf{S}$ are unitless in our notation). The first
term in the square brackets in Eq. (\ref{rene2}) is the isotropic hyperfine
Fermi-contact interaction while the second term is the dipole interaction.

\subsection{Interaction between electronic orbital motion and nuclear spin}

In Ref.\onlinecite{Dora}, it was shown that the orbital coupling does not
contribute to the frequency shift for a uniform state in zero magnetic
field. In this section, we show that there is a contribution from this
coupling for a non uniform state and in the presence of a transverse
magnetic field. Our starting point is the Hamiltonian\cite{Dora} of Eq. (\ref%
{couplingh}) for the coupling of the magnetic moment $\hslash \gamma _{n}%
\mathbf{I}$ of a nuclei at position $\mathbf{R}=0$ to the magnetic field $%
\mathbf{B}_{orb}\left( 0\right) $ due to the orbital motion of \textit{one}
electron. We then rewrite this Hamiltonian in second quantized form and
generalize it to nonuniform states.

Using Eq. (\ref{currentg}) for the electric current in graphene, the orbital
coupling can be written as 
\begin{equation}
h_{orb}\left( \mathbf{r}\right) =\xi C\frac{1}{r^{3}}(\tau _{x}y+\xi \tau
_{y}x),
\end{equation}%
where the constant%
\begin{equation}
C=-\frac{\mu _{0}}{4\pi }e\nu _{F}\hslash \gamma _{n}I_{z}.
\end{equation}

Assuming that Landau level mixing can be neglected, the second quantized
form of $h_{orb,\xi }\left( \mathbf{r}\right) $ for the interaction between
the magnetic moment of one nucleus at $\mathbf{R}_{m}$ and \textit{all}
conduction electrons in Landau level $n$ is given by%
\begin{eqnarray}
H_{orb,n}\left( \mathbf{R}_{m}\right) &=&\sum_{\xi ,\sigma }\int d\mathbf{r}%
\Psi _{\xi ,n,\sigma }^{\dagger }(\mathbf{r}) \\
&&\times h_{orb,\xi }\left( \mathbf{r-R}_{m}\right) \Psi _{\xi ,n,\sigma }(%
\mathbf{r}),  \notag
\end{eqnarray}%
where the field operator $\Psi _{\xi ,n,\sigma }(\mathbf{r})$ for an
electron in Landau level $n$ with valley index $\xi $ and with spin $\sigma $
is given by [we use the Landau gauge $\mathbf{A}=\left( 0,B_{0}x,0\right) $]

\begin{equation}
\Psi _{n,\xi ,\sigma }\left( \mathbf{r}\right) =\frac{1}{\sqrt{L_{y}}}%
\sum_{X}e^{-iXy/\ell ^{2}}\eta _{\xi ,n}\left( x-X\right) c_{\xi ,n,X,\sigma
},  \label{field}
\end{equation}%
where $n$ and $X$ are the Landau level and guiding-center indices
respectively and $L_{y}$ is the length of the 2DEG in the $y$ direction. The
spinors (in the sublattice basis) are given by 
\begin{eqnarray}
\eta _{-,n}\left( x\right) &=&\frac{1}{\sqrt{2}}\left( 
\begin{array}{c}
\mathrm{sgn}\left( n\right) i\varphi _{\left\vert n\right\vert -1}\left(
x\right) \\ 
\varphi _{\left\vert n\right\vert }\left( x\right)%
\end{array}%
\right) ,  \label{spinor} \\
\eta _{+,n}\left( x\right) &=&\frac{1}{\sqrt{2}}\left( 
\begin{array}{c}
\varphi _{\left\vert n\right\vert }\left( x\right) \\ 
-\mathrm{sgn}\left( n\right) i\varphi _{\left\vert n\right\vert -1}\left(
x\right)%
\end{array}%
\right) ,  \notag
\end{eqnarray}%
for $n>0.$ In these expressions, $\varphi _{n}\left( x\right) $ are the
eigenstates of the one-dimensional harmonic oscillator. For the special case 
$n=0,$ the spinors are%
\begin{eqnarray}
\eta _{-,0}\left( x\right) &=&\left( 
\begin{array}{c}
0 \\ 
\varphi _{0}\left( x\right)%
\end{array}%
\right) , \\
\eta _{+,0}\left( x\right) &=&\left( 
\begin{array}{c}
\varphi _{0}\left( x\right) \\ 
0%
\end{array}%
\right) .
\end{eqnarray}

For the orbital coupling, we then have%
\begin{eqnarray}
H_{orb,n}\left( \mathbf{R}_{m}\right) &=&\frac{1}{L_{y}}\sum_{X^{\prime
},X}\sum_{\xi ,\sigma }\int d\mathbf{r}e^{iX^{\prime }y/\ell
^{2}}e^{-iXy/\ell ^{2}}  \label{hdag} \\
&&\times D_{n,\xi ,X^{\prime },X}\left( \mathbf{r-R}_{m}\right) c_{\xi
,n,X^{\prime },\sigma }^{\dag }c_{\xi ,n,X,\sigma },  \notag
\end{eqnarray}%
where%
\begin{eqnarray}
D_{n,\xi ,X^{\prime },X}\left( \mathbf{r-R}_{m}\right) &=&\eta _{\xi
,n}^{\dag }\left( x-X^{\prime }\right) \\
&&\times h_{orb,\xi }\left( \mathbf{r-R}_{m}\right) \eta _{\xi ,n}\left(
x-X\right) .  \notag
\end{eqnarray}

We can use the operators $\rho _{\alpha ,\beta }\left( \mathbf{p}\right) $
defined by 
\begin{equation}
\rho _{\alpha ,\beta }\left( \mathbf{p}\right) =\frac{1}{N_{\varphi }}%
\sum_{X,X^{\prime }}e^{-\frac{i}{2}p_{x}\left( X+X^{\prime }\right) }\delta
_{X,X^{\prime }+p_{y}\ell ^{2}}c_{\alpha }^{\dagger }c_{\beta }
\label{orderp}
\end{equation}%
to write the product%
\begin{equation}
c_{\alpha }^{\dagger }c_{\beta }=\sum_{\mathbf{p}}\rho _{\alpha ,\beta
}\left( \mathbf{p}\right) e^{\frac{i}{2}p_{x}\left( X+X^{\prime }\right)
}\delta _{X^{\prime },X+p_{y}\ell ^{2}}.  \label{cdag}
\end{equation}%
The relation between the operators $\rho _{\alpha ,\beta }\left( \mathbf{p}%
\right) $ and the field operators is%
\begin{eqnarray}
&&\int d\mathbf{r\,}\Psi _{n,\xi ,\sigma }^{\dagger }\mathbf{\left( \mathbf{r%
}\right) }e^{-i\mathbf{p}\cdot \mathbf{r}}\Psi _{n^{\prime },\xi ^{\prime
},\sigma ^{\prime }}\left( \mathbf{r}\right) \\
&=&N_{\varphi }F_{n,n^{\prime }}\left( -\mathbf{p}\right) \rho _{n,\xi
,\sigma ;n^{\prime },\xi ^{\prime },\sigma ^{\prime }}\left( \mathbf{p}%
\right) ,  \notag
\end{eqnarray}%
where the function 
\begin{eqnarray}
F_{n,n^{\prime }}\left( \mathbf{p}\right) &=&\sqrt{\frac{Min\left(
n,n^{\prime }\right) !}{Max\left( n,n^{\prime }\right) !}}\left( \frac{%
\left( \pm p_{y}+ip_{x}\right) \ell }{\sqrt{2}}\right) ^{\left\vert
n-n^{\prime }\right\vert }  \label{fnn} \\
&&\times L_{Min\left( n,n^{\prime }\right) }^{\left\vert n-n^{\prime
}\right\vert }\left( \frac{p^{2}\ell ^{2}}{2}\right) e^{-\frac{p^{2}\ell ^{2}%
}{4}},  \notag
\end{eqnarray}%
with $L_{n}^{m}\left( x\right) $ the generalized Laguerre polynomials. The
average values $\left\langle \rho _{\alpha ,\beta }\left( \mathbf{p}\right)
\right\rangle $ can be interpreted as the order parameters of the crystal
state. Inserting Eq. (\ref{cdag}) in Eq. (\ref{hdag}), we get%
\begin{eqnarray}
H_{orb,n}\left( \mathbf{R}_{m}\right) &=&\frac{1}{L_{y}}\sum_{\mathbf{p}%
}\sum_{\xi ,\sigma }\sum_{X,X^{\prime }}\int d\mathbf{r}e^{-i\left(
X-X^{\prime }\right) y/\ell ^{2}}  \label{horbi} \\
&&\times e^{\frac{i}{2}p_{x}\left( X+X^{\prime }\right) }\delta _{X^{\prime
},X+p_{y}\ell ^{2}}  \notag \\
&&\times \rho _{\xi ,n,\sigma ;\xi ,n,\sigma }\left( \mathbf{p}\right)
D_{n,\xi ,X^{\prime },X}\left( \mathbf{r-R}_{m}\right) .  \notag
\end{eqnarray}%
Because of the particular form of the spinors in Landau level $n=0,$ we have
immediately $H_{orb,n=0}=0.$ Thus, we need only evaluate the last term in
Eq. (\ref{horbi}) for $n\neq 0.$ We get%
\begin{eqnarray}
D_{n,\xi ,X^{\prime },X}\left( \mathbf{r-R}_{m}\right) &=&\frac{i}{2}\mathrm{%
sgn}\left( n\right) C\frac{\Lambda \left( \mathbf{r}-\mathbf{R}_{m}\right) }{%
\left\vert \mathbf{r}-\mathbf{R}_{m}\right\vert ^{3}} \\
&&\times \left[ \Lambda \left( \mathbf{r}-\mathbf{R}_{m}\right) \Phi
_{\left\vert n\right\vert -1,X^{\prime };\left\vert n\right\vert ,X}\left(
x\right) \right.  \notag \\
&&\left. -\Lambda ^{\ast }\left( \mathbf{r}-\mathbf{R}_{m}\right) \Phi
_{\left\vert n\right\vert ,X^{\prime };\left\vert n\right\vert -1,X}\left(
x\right) \right] ,  \notag
\end{eqnarray}%
where the function%
\begin{equation}
\Lambda \left( \mathbf{r}-\mathbf{R}_{m}\right) =\left( y+ix\right) -\left(
y_{m}+ix_{m}\right)
\end{equation}%
with $\mathbf{R}_{m}=\left( x_{m},y_{m}\right) $ and 
\begin{equation}
\Phi _{nX^{\prime };mX}\left( x\right) =\varphi _{n}\left( x-X^{\prime
}\right) \varphi _{m}\left( x-X\right) .
\end{equation}

We define the Fourier transform%
\begin{equation}
\varphi _{n}\left( x-X\right) =\frac{1}{L_{x}}\sum_{t}e^{it\left( x-X\right)
}\widetilde{\varphi }_{n}\left( t\right) ,
\end{equation}%
so that we can write%
\begin{eqnarray}
&&\sum_{X^{\prime },X}e^{-i\left( X-X^{\prime }\right) y/\ell ^{2}}e^{\frac{i%
}{2}p_{x}\left( X+X^{\prime }\right) } \\
&&\times \delta _{X^{\prime },X+p_{y}\ell ^{2}}\Phi _{\left\vert
n\right\vert ,X^{\prime },\left\vert m\right\vert ,X}\left( x\right)  \notag
\\
&=&\frac{N_{\varphi }}{L_{x}^{2}}e^{\frac{i}{2}p_{x}p_{y}\ell ^{2}}e^{i%
\mathbf{p}\cdot \mathbf{r}}  \notag \\
&&\times \sum_{t}\widetilde{\varphi }_{\left\vert n\right\vert }\left(
t\right) \widetilde{\varphi }_{\left\vert m\right\vert }\left(
p_{x}-t\right) e^{-itp_{y}\ell ^{2}},  \notag
\end{eqnarray}%
using the identity%
\begin{equation}
\frac{1}{N_{\varphi }}\sum_{X}e^{iX\left( p-q\right) }=\delta _{p,q}.
\end{equation}

With the change of variables $\mathbf{r}\rightarrow \mathbf{r}+\mathbf{R}%
_{m},$ we get%
\begin{eqnarray}
H_{orb,n} &=&-i\mathrm{sgn}\left( n\right) C\frac{1}{2L_{y}}\frac{N_{\varphi
}}{L_{x}^{2}}\sum_{\xi ,\sigma }\sum_{\mathbf{p}}\sum_{t} \\
&&\times \int d\mathbf{r}e^{i\mathbf{p}\cdot \mathbf{r}}\rho _{\xi ,n,\sigma
;\xi ,n,\sigma }\left( \mathbf{p}\right) e^{\frac{i}{2}p_{x}p_{y}\ell
^{2}}e^{-itp_{y}\ell ^{2}}  \notag \\
&&\times \frac{e^{i\mathbf{p}\cdot \mathbf{R}_{m}}}{\left\vert \mathbf{r}-%
\mathbf{R}_{m}\right\vert ^{3}}\left[ -\Lambda \left( \mathbf{r}\right) 
\widetilde{\varphi }_{\left\vert n\right\vert -1}\left( t\right) \widetilde{%
\varphi }_{\left\vert n\right\vert }\left( p_{x}-t\right) \right.  \notag \\
&&\left. +\Lambda ^{\ast }\left( \mathbf{r}\right) \widetilde{\varphi }%
_{\left\vert n\right\vert }\left( t\right) \widetilde{\varphi }_{\left\vert
n\right\vert -1}\left( p_{x}-t\right) \right] ,  \notag
\end{eqnarray}%
where $\Lambda \left( \mathbf{r}\right) $ has now been redefined as\qquad 
\begin{equation}
\Lambda \left( \mathbf{r}\right) =y-\xi ix.
\end{equation}

Another change of variables: $t\rightarrow t+p_{x}$ in the second line of $%
H_{orb,\xi ,n}$ gives 
\begin{eqnarray}
H_{orb,n} &=&\mathrm{sgn}\left( n\right) \frac{\mu _{0}}{4\pi }ie^{2}\nu
_{F}B_{0}\gamma _{n}I_{z} \\
&&\times \sum_{\xi ,\sigma }\sum_{\mathbf{p}}e^{i\mathbf{p}\cdot \mathbf{R}%
_{m}}\rho _{\xi ,n,\sigma ;\xi ,n,\sigma }\left( \mathbf{p}\right) \Upsilon
_{\left\vert n\right\vert >0}\left( \mathbf{p}\right) ,  \notag
\end{eqnarray}%
where%
\begin{eqnarray}
\Upsilon _{\left\vert n\right\vert >0}\left( \mathbf{p}\right) &=&\frac{1}{%
L_{x}}\sum_{t}\widetilde{\varphi }_{\left\vert n\right\vert -1}\left(
t\right) \widetilde{\varphi }_{\left\vert n\right\vert }\left( p_{x}-t\right)
\\
&&\times \cos \left( \frac{p_{x}p_{y}\ell ^{2}}{2}-tp_{y}\ell ^{2}-\theta _{%
\mathbf{p}}\right)  \notag
\end{eqnarray}%
and $\theta _{\mathbf{p}}$ is the angle between the vector $\mathbf{p}$ and
the $x$ axis.

The function $\Upsilon _{\left\vert n\right\vert >0}\left( \mathbf{p}\right) 
$ can be written as%
\begin{eqnarray}
\Upsilon _{\left\vert n\right\vert }\left( \mathbf{p}\right) &=&\frac{1}{2}%
e^{-i\left( \frac{p_{x}p_{y}\ell ^{2}}{2}+\theta _{\mathbf{p}}\right)
}G_{\left\vert n\right\vert -1,\left\vert n\right\vert }\left(
p_{x},p_{y}\right) \\
&&+\frac{1}{2}e^{i\left( \frac{p_{x}p_{y}\ell ^{2}}{2}+\theta _{\mathbf{p}%
}\right) }G_{\left\vert n\right\vert -1,\left\vert n\right\vert }\left(
p_{x},-p_{y}\right) ,  \notag
\end{eqnarray}%
where%
\begin{equation}
G_{n,m}\left( p_{x},p_{y}\right) =\int_{-\infty }^{\infty }dx\varphi
_{\left\vert n\right\vert -1}\left( x\right) \varphi _{\left\vert
n\right\vert }\left( x+p_{y}\ell ^{2}\right) e^{-ip_{x}x}.
\end{equation}

Using the definitions of the normalized eigenfunctions for the
one-dimensional harmonic oscillator:%
\begin{eqnarray}
\varphi _{0}\left( x\right) &=&\left( \frac{1}{\pi \ell ^{2}}\right)
^{1/4}e^{-x^{2}/2\ell ^{2}}, \\
\varphi _{1}\left( x\right) &=&\left( \frac{1}{\pi \ell ^{2}}\right)
^{1/4}\left( \sqrt{2}\frac{x}{\ell }\right) e^{-x^{2}/2\ell ^{2}}, \\
\varphi _{2}\left( x\right) &=&\left( \frac{1}{\pi \ell ^{2}}\right) ^{1/4}%
\frac{1}{\sqrt{2}}\left( 2\left( \frac{x}{\ell }\right) ^{2}-1\right)
e^{-x^{2}/2\ell ^{2}},
\end{eqnarray}%
we get for the functions $\Upsilon _{n}$ 
\begin{eqnarray}
\Upsilon _{n=0}\left( \mathbf{p}\right) &=&0, \\
\Upsilon _{n=1}\left( \mathbf{p}\right) &=&-\frac{1}{\sqrt{2}}ip\ell e^{-%
\frac{1}{4}p^{2}\ell ^{2}}, \\
\Upsilon _{n=2}\left( \mathbf{p}\right) &=&\frac{1}{4}ie^{-\frac{1}{4}%
p^{2}\ell ^{2}}p\ell \left( p^{2}\ell ^{2}-4\right) .
\end{eqnarray}%
In a uniform state, the average $\left\langle \rho _{\xi ,n,\sigma ;\xi
,n,\sigma }\left( \mathbf{p}\right) \right\rangle \neq 0$ for $\mathbf{p}=0$
only and $\Upsilon _{\left\vert n\right\vert >0}\left( \mathbf{p}=0\right)
=0 $ so that the orbital shift vanishes in this case.

The orbital coupling can be written in terms of an effective magnetic field $%
B_{z,eff}$ defined by%
\begin{equation}
\left\langle H_{orb,n}\left( \mathbf{R}_{m}\right) \right\rangle \equiv
-\hslash \gamma _{n}I_{z}B_{z,eff,n}.  \label{rene13}
\end{equation}%
This additional magnetic field leads to a frequency shift 
\begin{eqnarray}
\Delta f_{orb,n}\left( \mathbf{R}_{m}\right) &=&\frac{\gamma _{n}}{2\pi }%
B_{z,eff,n}  \label{rene15} \\
&=&-i\alpha _{orb}B_{0}\mathrm{sgn}\left( n\right)  \notag \\
&&\times \sum_{\xi ,\sigma }\sum_{\mathbf{p}}e^{i\mathbf{p}\cdot \mathbf{R}%
_{m}}\Upsilon _{\left\vert n\right\vert }\left( \mathbf{p}\right)
\left\langle \rho _{\xi ,n,\sigma ;\xi ,n,\sigma }\left( \mathbf{p}\right)
\right\rangle ,  \notag
\end{eqnarray}%
where the constant%
\begin{equation}
\alpha _{orb}=\frac{e^{2}}{h}\frac{\mu _{0}}{4\pi }\gamma _{n}\nu _{F}=0.236%
\text{ kHz/T.}
\end{equation}

We remark that the filled levels below Landau level $n$ do not contribute to
the orbital shift since a filled state implies a uniform electronic density
and $\Upsilon _{n}\left( \mathbf{p}=0\right) =0$. Also $\Delta f_{orb,n}$
depends on the sign of $n$ as expected since the current can be viewed as
assumed by the holes for $n<0$ and by the electrons for $n>0.$

\subsection{Interaction between the electronic spin and the nuclear spin}

The coupling between the electronic spin and the spin of an ion at $\mathbf{R%
}_{m}=0$ is given by%
\begin{equation}
h_{spin}\left( \mathbf{r}\right) =h_{iso}\left( \mathbf{r}\right)
+h_{dip}\left( \mathbf{r}\right) ,
\end{equation}%
where 
\begin{equation}
h_{iso}\left( \mathbf{r}\right) =\frac{2}{3}\mu _{0}g\mu _{B}\hslash \gamma
_{n}\delta \left( \mathbf{r}\right) \mathbf{I}\cdot \mathbf{S}_{\xi }
\end{equation}%
is the isotropic Fermi-contact interaction and the dipole contribution is
given by%
\begin{equation}
h_{dip}\left( \mathbf{r}\right) =-\frac{\mu _{0}}{4\pi }g\mu _{B}\hslash
\gamma _{n}\mathbf{S}_{\xi }\mathbf{\cdot a}^{\mathrm{dip}}\left( \mathbf{r}%
\right) \cdot \mathbf{I.}  \label{hdip}
\end{equation}%
The term $\mathbf{a}^{\mathrm{dip}}$ in Eq. (\ref{hdip}) is a traceless
matrix given, using Eq. (\ref{rene2}) (with $\mathbf{r}$ a 3d vector in
these two equations), by 
\begin{equation}
a_{i,j}^{\mathrm{dip}}\left( \mathbf{r}\right) =\frac{1}{r^{3}}\delta _{i,j}-%
\frac{3}{r^{5}}r_{i}r_{j},
\end{equation}%
where $i=x,y,z$.

The many-body spin coupling between all electrons in Landau level $n$ and
one ion at $\mathbf{R}_{m}$ is 
\begin{eqnarray}
H_{spin,n}\left( \mathbf{R}_{m}\right) &=&\sum_{\xi ,\sigma }\int d\mathbf{r}%
\Psi _{\xi ,n,\sigma }^{\dagger }(\mathbf{r}) \\
&&\left. \times h_{spin}\left( \mathbf{r-R}_{m}\right) \Psi _{\xi ,n,\sigma
}(\mathbf{r})\right\rangle .  \notag
\end{eqnarray}%
The coupling $H_{spin,n}$ is sensitive to the microscopic electronic spin
density near each nucleus, something that is not captured by the continuum
approximation. For example, the spin coupling depends on the periodic part $%
u_{n,\mathbf{k}}\left( \mathbf{r}\right) $ of the Bloch function evaluated
at the position of the nucleus\cite{Slichter}. Thus, the spin coupling must
be obtained from first principle calculations. One such calculation is done
in Ref. \onlinecite{Yazyev} for carbon nanotubes and graphene fragments. In
this reference, a local axial symmetry is assumed so that the averaged $%
a_{xx}^{\mathrm{dip}}=a_{yy}^{\mathrm{dip}}=-\frac{1}{2}a_{zz}^{\mathrm{dip}%
} $ and the other averaged components of the $\mathbf{a}^{\mathrm{dip}}$
matrix are zero. If the nuclear spins are assumed completely polarized along
the external magnetic field, then 
\begin{eqnarray}
\left\langle H_{spin,n}\left( \mathbf{R}_{m}\right) \right\rangle &=&h\left(
A^{\mathrm{iso}}+A_{zz}^{\mathrm{dip}}\right) I_{z}\left\langle
S_{z,n}\left( \mathbf{R}_{m}\right) \right\rangle  \label{rene5} \\
&=&hA_{zz}^{\mathrm{spin}}I_{z}\left\langle S_{z,n}\left( \mathbf{R}%
_{m}\right) \right\rangle ,  \notag
\end{eqnarray}%
where $\left\langle S_{z,n}\left( \mathbf{R}_{m}\right) \right\rangle $ is
the average effective electronic spin around each carbon atom from electrons
in Landau level $n$ (including both valleys) and the nuclear spin $I_{z}=1/2$%
.

In our numerical calculations, we use the values of $A^{\mathrm{iso}}$ and $%
A_{zz}^{\mathrm{dip}}$ quoted in Ref. \onlinecite{Dora} which are based on
the calculations of Ref. \onlinecite{Yazyev}. These values are $A_{zz}^{%
\mathrm{dip}}=146$ MHz and $A^{\mathrm{iso}}=-44$ MHz so that $A_{zz}^{%
\mathrm{spin}}=102$ MHz.

For a Wigner crystal, the spin polarization varies spatially and is much
less that $S_{z}=1/2$ around a single carbon atom. The unit cell area of the
graphene lattice is $A_{c}=3\sqrt{3}c^{2}/2=5.\,\allowbreak 24\times
10^{-20} $ m$^{2}$. The area occupied by one carbon atom is $A_{c}/2$ and so 
$S_{z,n}$ in Eq. (\ref{rene5}) is given by

\begin{equation}
\left\langle S_{z,n}\left( \mathbf{r}\right) \right\rangle =\frac{A_{c}}{4}%
\sum_{\xi }\sum_{\sigma ,\sigma ^{\prime }}\left\langle \Psi _{n,\xi ,\sigma
}^{\dag }\left( \mathbf{r}\right) \sigma _{\sigma \sigma ^{\prime }}^{\left(
z\right) }\Psi _{n,\xi ,\sigma ^{\prime }}\left( \mathbf{r}\right)
\right\rangle ,
\end{equation}%
or, for an ion at $\mathbf{R}_{m},$%
\begin{equation}
\left\langle S_{z,n}\left( \mathbf{R}_{m}\right) \right\rangle =\frac{1}{%
2\pi \ell ^{2}}\frac{1}{2}A_{c}\sum_{\mathbf{G}}e^{i\mathbf{G}\cdot \mathbf{R%
}_{m}}F_{n,n}\left( -\mathbf{G}\right) \left\langle \rho _{z,n}\left( 
\mathbf{G}\right) \right\rangle ,
\end{equation}%
where 
\begin{equation}
\left\langle \rho _{z,n}\left( \mathbf{G}\right) \right\rangle =\frac{1}{2}%
\sum_{\xi }\left[ \left\langle \rho _{n,\xi ,+;n,\xi ,+}\left( \mathbf{G}%
\right) \right\rangle -\left\langle \rho _{n,\xi ,-;n,\xi ,-}\left( \mathbf{G%
}\right) \right\rangle \right]
\end{equation}%
and $\mathbf{G}$ is a reciprocal lattice vector of the crystal.

The spin coupling is finally given by%
\begin{eqnarray}
\left\langle H_{spin,n}\left( \mathbf{R}_{m}\right) \right\rangle
&=&hI_{z}A_{zz}^{\mathrm{spin}}\frac{A_{c}}{4\pi \ell ^{2}}\sum_{\mathbf{G}%
}e^{i\mathbf{G}\cdot \mathbf{R}_{m}} \\
&&\times F_{n,n}\left( -\mathbf{G}\right) \left\langle \rho _{z,n}\left( 
\mathbf{G}\right) \right\rangle  \notag
\end{eqnarray}%
where the functions%
\begin{eqnarray}
F_{0,0}\left( \mathbf{G}\right) &=&e^{-G^{2}\ell ^{2}/4}, \\
F_{1,1}\left( \mathbf{G}\right) &=&\left( 1-\frac{G^{2}\ell ^{2}}{2}\right)
e^{-G^{2}\ell ^{2}/4}, \\
F_{2,2}\left( \mathbf{G}\right) &=&\left( 1-G^{2}\ell ^{2}+\frac{G^{4}\ell
^{4}}{8}\right) e^{-G^{2}\ell ^{2}/4}.
\end{eqnarray}

Using Eq. (\ref{rene13}) to define an effective magnetic field for the spin
coupling, we arrive at the following result for the Knight shift at the
position of a nuclei: 
\begin{equation}
\Delta f_{spin}\left( \mathbf{R}_{m}\right) =-\alpha _{spin}B_{0}\sum_{%
\mathbf{G}}e^{i\mathbf{G}\cdot \mathbf{R}_{m}}F_{n,n}\left( -\mathbf{G}%
\right) \left\langle \rho _{z,n}\left( \mathbf{G}\right) \right\rangle
\end{equation}%
with the constant%
\begin{equation}
\alpha _{spin}=A_{zz}^{\mathrm{spin}}\frac{eA_{c}}{2h}=0.646\text{ kHz/T.}
\label{alpha}
\end{equation}

For graphene, the ground state of the uniform 2DEG has $\left\langle \rho
_{z,n}\left( 0\right) \right\rangle =-1$ for $\nu _{n}\in \left[ 0,2\right]
, $ where $\nu _{n}$ is the filling factor of Landau level $n,$ so that the
shift $\Delta f_{spin}$ is positive when $\alpha _{spin}$ is given by Eq. (%
\ref{alpha}).

\section{ORDER PARAMETERS OF THE CRYSTAL STATES}

We summarize in this section the calculation of the order parameters $%
\left\langle \rho _{\beta ,\alpha }\left( \mathbf{G}\right) \right\rangle $
defined in Eq. (\ref{orderp}) for the crystal phases. The formalism for this
calculation is described in detail in Ref. \onlinecite{CoteSkyrmeGraphene}
for the case where the spin degree of freedom is neglected. In this paper,
we include both valley and spin states. To simplify the notation, we use $%
\alpha ,\beta $ for the combined valley and spin indices. All quantities are
evaluated in Landau level $n$ and so we leave this index implicit hereafter.

The order parameters are related to the single-particle Matsubara Green's
function by%
\begin{equation}
\left\langle \rho _{\beta ,\alpha }\left( \mathbf{G}\right) \right\rangle
=G_{\alpha ,\beta }\left( \mathbf{G,}\tau =0^{-}\right) ,
\end{equation}%
where the Fourier transform of the Green's function is given by 
\begin{eqnarray}
G_{\alpha ,\beta }\left( \mathbf{G,}\tau \right) &=&\frac{1}{N_{\varphi }}%
\sum_{X,X^{\prime }}e^{-\frac{i}{2}G_{x}\left( X+X^{\prime }\right) } \\
&&\times \delta _{X,X^{\prime }-G_{y}\ell ^{2}}G_{\alpha ,\beta }\left(
X,X^{\prime },\tau \right)  \notag
\end{eqnarray}%
with the Green's function defined by 
\begin{equation}
G_{\alpha ,\beta }\left( X,X^{\prime },\tau \right) =-\left\langle T_{\tau
}c_{X,\alpha }\left( \tau \right) c_{X^{\prime },\beta }^{\dagger }\left(
0\right) \right\rangle ,
\end{equation}%
with $T_{\tau }$ the time-ordering operator and $\tau $ the imaginary time.

The Green's function is evaluated in the Hartree-Fock approximation with an
Hamiltonian given by (with $\kappa =4\pi \varepsilon _{r}\varepsilon _{0}$) 
\begin{eqnarray}
H &=&N_{\varphi }\sum_{\alpha }E_{\alpha }\rho _{\alpha ,\alpha }\left( 
\mathbf{G}=0\right) \\
&&+N_{\varphi }\left( \frac{e^{2}}{\kappa \ell }\right) \sum_{\alpha ,\beta
}\sum_{\mathbf{G}}H_{n}\left( \mathbf{G}\right) \left\langle \rho _{\alpha
,\alpha }\left( \mathbf{G}\right) \right\rangle \rho _{\beta ,\beta }\left( -%
\mathbf{G}\right)  \notag \\
&&-N_{\varphi }\left( \frac{e^{2}}{\kappa \ell }\right) \sum_{\alpha ,\beta
}\sum_{\mathbf{G}}X_{n}\left( \mathbf{G}\right) \left\vert \left\langle \rho
_{\alpha ,\beta }\left( \mathbf{G}\right) \right\rangle \right\vert ^{2}. 
\notag
\end{eqnarray}%
The term $E_{\alpha }=\sigma g\mu _{B}B/2$ is simply the Zeeman energy. The
equation of motion for $G_{\alpha ,\beta }\left( \mathbf{G,}i\omega
_{n}\right) ,$ where $\omega _{n}$ are the Matsubara frequencies, is%
\begin{eqnarray}
&&\left[ i\omega _{n}-\left( E_{\alpha }-\mu \right) /\hslash \right]
G_{\alpha ,\beta }\left( \mathbf{G},\omega _{n}\right) =\delta _{\mathbf{G}%
,0}\delta _{\alpha ,\beta }  \label{rene14} \\
&&+\frac{1}{\hslash }\sum_{\mathbf{G}^{\prime }\neq \mathbf{G}}U^{H}\left( 
\mathbf{G-G}^{\prime }\right) e^{-i\left( \mathbf{G}\times \mathbf{G}%
^{\prime }\right) \cdot \widehat{\mathbf{z}}\ell ^{2}/2}G_{\alpha ,\beta
}\left( \mathbf{G}^{\prime },\omega _{n}\right)  \notag \\
&&-\frac{1}{\hslash }\sum_{\mathbf{G}^{\prime }}\sum_{\gamma }U_{\alpha
,\gamma }^{F}\left( \mathbf{G-G}^{\prime }\right) e^{-i\left( \mathbf{G}%
\times \mathbf{G}^{\prime }\right) \cdot \widehat{\mathbf{z}}\ell
^{2}/2}G_{\gamma ,\beta }\left( \mathbf{G}^{\prime },\omega _{n}\right) , 
\notag
\end{eqnarray}%
with the potentials%
\begin{eqnarray}
U^{H}\left( \mathbf{G}\right) &=&\left( \frac{e^{2}}{\kappa \ell }\right)
\sum_{\gamma }H_{n}\left( -\mathbf{G}\right) \left\langle \rho _{\gamma
,\gamma }\left( -\mathbf{G}\right) \right\rangle , \\
U_{\alpha ,\beta }^{F}\left( \mathbf{G}\right) &=&\left( \frac{e^{2}}{\kappa
\ell }\right) X_{n}\left( -\mathbf{G}\right) \left\langle \rho _{\beta
,\alpha }\left( -\mathbf{G}\right) \right\rangle ,
\end{eqnarray}%
and the Hartree and Fock interactions in Landau level $n$ given by%
\begin{eqnarray}
H_{n}\left( \mathbf{G}\right) &=&\frac{1}{G\ell }\Xi _{n}^{2}\left( \mathbf{G%
}\right) ,  \label{fock} \\
X_{n}\left( \mathbf{G}\right) &=&\frac{2\pi \ell ^{2}}{S}\sum_{\mathbf{p}%
}H_{n}\left( \mathbf{p}\right) e^{i\mathbf{p\times G}\ell ^{2}},  \notag
\end{eqnarray}%
where the form factor%
\begin{eqnarray}
\Xi _{n}\left( \mathbf{G}\right) &=&\delta _{n,0}F_{0,0}\left( \mathbf{G}%
\right) \\
&&+\frac{1}{2}\Theta \left( \left\vert n\right\vert \right) \left[
F_{\left\vert n\right\vert ,\left\vert n\right\vert }\left( \mathbf{G}%
\right) +F_{\left\vert n\right\vert -1,\left\vert n\right\vert -1}\left( 
\mathbf{G}\right) \right] .  \notag
\end{eqnarray}

Eq. (\ref{rene14}) is solved iteratively until convergence is reached for
the order parameters.

\section{NMR\ SPECTRA OF SOME\ CRYSTAL PHASES}

Because of the finite Zeeman coupling$,$ the noninteracting uniform 2DEG in
graphene is fully spin polarized for filling factor $\nu _{n}\in \left[ 0,2%
\right] $ and partially spin polarized for $\nu _{n}\in \left[ 2,4\right] .$
Levels of same spin but different valley indices are degenerate. In
graphene, the electron spin and magnetic moments are antiparallel so that
the two lower noninteracting energy levels have spin $\sigma =-1.$

In this paper, we use a pseudospin language where the two valley states $%
K_{+},K_{-}$ are mapped onto the up and down valley pseudospin states
respectively$.$ It follows that the pseudospin vector field in real space is
obtained from 
\begin{equation}
\mathbf{P}\left( \mathbf{r}\right) =\frac{1}{2\pi \ell ^{2}}\sum_{\mathbf{G}%
}e^{i\mathbf{G}\cdot r}F_{n,n}\left( -\mathbf{G}\right) \left\langle \mathbf{%
p}\left( \mathbf{G}\right) \right\rangle ,
\end{equation}%
where the function $F_{n,n}\left( \mathbf{G}\right) $ is defined in Eq. (\ref%
{fnn}) and the Cartesian components of $\mathbf{P}\left( \mathbf{r}\right) $
are given by 
\begin{eqnarray}
p_{x}\left( \mathbf{G}\right) &=&\frac{1}{2}\sum_{\sigma }\left[ \rho
_{+,\sigma :-,\sigma }\left( \mathbf{G}\right) +\rho _{-,\sigma :+,\sigma
}\left( \mathbf{G}\right) \right] , \\
p_{y}\left( \mathbf{G}\right) &=&\frac{1}{2i}\sum_{\sigma }\left[ \rho
_{+,\sigma :-,\sigma }\left( \mathbf{G}\right) -\rho _{-,\sigma :+,\sigma
}\left( \mathbf{G}\right) \right] , \\
p_{z}\left( \mathbf{G}\right) &=&\frac{1}{2}\sum_{\sigma }\left[ \rho
_{+,\sigma :+,\sigma }\left( \mathbf{G}\right) -\rho _{-,\sigma :-,\sigma
}\left( \mathbf{G}\right) \right] .
\end{eqnarray}%
For the spin field in real space, we have in the same way%
\begin{eqnarray}
s_{x}\left( \mathbf{G}\right) &=&\frac{1}{2}\sum_{\xi }\left[ \rho _{\xi
,+:\xi ,-}\left( \mathbf{G}\right) +\rho _{\xi ,-:\xi ,+}\left( \mathbf{G}%
\right) \right] , \\
s_{y}\left( \mathbf{G}\right) &=&\frac{1}{2}\sum_{\xi }\left[ \rho _{\xi
,+:\xi ,-}\left( \mathbf{G}\right) -\rho _{\xi ,-:\xi ,+}\left( \mathbf{G}%
\right) \right] , \\
s_{z}\left( \mathbf{G}\right) &=&\frac{1}{2}\sum_{\xi }\left[ \rho _{\xi
,+:\xi ,+}\left( \mathbf{G}\right) -\rho _{\xi ,-:\xi ,-}\left( \mathbf{G}%
\right) \right] .
\end{eqnarray}%
Finally, for the electronic density 
\begin{equation}
n\left( \mathbf{G}\right) =\sum_{\xi ,\sigma }\rho _{\xi ,\sigma :\xi
,\sigma }\left( \mathbf{G}\right) .
\end{equation}

In the Hartree-Fock approximation, the electrons condense into distinct
quantum Hall ferromagnetic states at filling factors $\nu _{n}=1,2,3,4$ and
the (approximate) fourfold degeneracy is fully lifted. The ground state at
integer filling is found\cite{Doretto} by (1)\ maximizing the spin and then
(2) maximizing the valley pseudospin $\left\vert \mathbf{P}\right\vert $ to
the extent allowed by rule (1). Thus, for the uniform 2DEG, the lower(upper)
energy levels have spin per electron $s_{z}=-1/2$($s_{z}=+1/2$). Since there
is no effective Zeeman coupling for the valley pseudospin $\mathbf{P}$, this
vector has a full SU(2) symmetry. For $\nu =0$ (i.e., $\nu _{0}=2$),
Hartree-Fock does not explain the experimental observations and a more
sophisticated analysis that includes the electron-phonon interaction is
needed\cite{Kharitonov}.

In the 2DEG, the Coulomb interaction in the presence of a quantizing
magnetic field can also favor the formation of crystal states such as the
Wigner crystal\cite{Wigner}. These states have been studied in a number of
papers for graphene\cite{CoteSkyrmeGraphene,GoerbigCW,Joglekar} as well as
for multilayer graphene\cite{Yoshioka}. Several crystalline ground states
were identified in Landau levels $n=0,1,2,3:\ $triangular Wigner and bubble
crystals\cite{Revuebulles,Cotebubble} of electrons or holes and square
crystals with a meronlike valley pseudospin texture. The spin degree of
freedom was not considered in these previous works so that all these
crystalline phases are fully spin polarized. In the present paper, we
explicitly consider valley and spin degrees of freedom. For small values of
the Zeeman coupling, this allows for the addition of a new phase in the
phase diagram which is a crystal of spin skyrmions.

To compute the NMR\ line shape $I\left( f\right) $ of the crystal phases, we
need a value for the linewidth $\Gamma $ in Eq. (\ref{gamma}) i.e., the
linewidth of the NMR line shape for a uniform state of the 2DEG in a $^{13}$%
C graphene sheet. Although there is a number of theoretical calculations of
NMR parameters in carbon nanotubes and graphene\cite{Fabio}, there are at
present no available NMR\ experimental results for a single graphene layer
to validate them. If we assume a linewidth limited only by the nuclear
dipole-dipole interaction, then, for nuclei that are separated by a distance 
$r$ and have magnetic moment $\mu =\hslash \gamma _{n}I,$ a rough estimate%
\cite{Slichter} of the local magnetic field they produce is $B_{loc}=\mu
/r^{3}$ in cgs units. Since this local field can increase or decrease the
external field, the spread in frequency that results is $\Delta f=2\left(
\gamma _{n}/2\pi \right) B_{loc}$ for the FWHM. For graphene, this leads to
a value $\Gamma \approx 1$ kHz independent of the magnetic field. Our
strategy is to start with this value of $\Gamma $ and increase it until we
find the maximal value above which there is no difference between the line
shape of the crystal and that of the uniform state with the same average
filling factor.

The intensity of the signal is of course of major importance experimentally.
This issue was discussed in Ref. \onlinecite{Dora}. In this paper, the
authors estimated the limit of detection (LOD) parameter for a fully $^{13}$%
C enriched graphene layer using the results available for enriched
fullerenes and carbon nanotubes. They concluded that, with the improvement
in the synthesis of a fully $^{13}$C enriched graphene, NMR detection should
be within reach. Assuming this to be the case, we will not concern ourselves
with this problem in this paper.

In all the NMR\ line shapes shown in the figures below, the frequency is
measured with respect to the resonance frequency $f_{0}.$ For comparison, we
give the line shape of the crystal and that of the uniform state with the
same filling factor. For a uniform state (US), the line shape is 
\begin{equation}
I_{US}\left( f\right) =e^{-\left( f-\Delta f_{spin}\right) ^{2}/\Gamma ^{2}}
\end{equation}%
since $\Delta f_{orb}=0$ when there is no density modulation. Also, in such
a state, $\Delta f_{spin}=-\alpha _{spin}B_{0}\left\langle \rho _{z,n}\left(
0\right) \right\rangle $ is directly proportional to the spin polarization.
In our calculations, we first set the Zeeman coupling $\Delta _{Z}.$ This,
in turn, determines the value of the magnetic field $B_{0}$ and the
electronic density when the filling factor is specified.

Figure 1 shows the NMR\ line shape of different Wigner crystals obtained
with the Zeeman coupling (expressed in units of $\left( e^{2}/\kappa \ell
\right) $) $\Delta _{Z}\equiv g\mu _{B}B/\left( e^{2}/\kappa \ell \right)
=0.03.$ Figure 1(a) shows how the line shape changes with Landau level index 
$n=0,1,2$ for $\nu _{n}=0.2$ where the ground state (in the Hartree-Fock
approximation) is an electron Wigner crystal. Note that the line shape of
the uniform state is the same for all three crystals. Also, there is no
orbital shift for $n=0$ while the signal for the other two Landau levels is
a combination of spin and orbital shifts. For $n=0,$ the line shape is
similar to that of the $n=0$ Wigner crystal in GaAs quantum wells obtained
from RD-NMR. This is expected because the form factor function $%
F_{0,0}\left( \mathbf{G}\right) $ is the same in graphene and GaAs quantum
wells when $n=0$ and the orbital shift is negligible in GaAs\cite%
{CoteNMR,Tiemann}. Figure 1 shows that the line shape changes qualitatively
with Landau level index. In all cases shown in Fig. 1(a), the signal of the
crystal is clearly distinguishable from that of the uniform state. This is
also true for the other types of Wigner crystals found in graphene. For
example, Fig. 1(b) shows the line shapes, in level $n=2,$ of the Wigner
crystal at $\nu _{2}=0.2$ and of the bubble crystal (a Wigner crystal with
two electrons per site) at $\nu _{2}=0.4$ while Fig. 1(c) shows the line
shape, in level $n=1,$ of the Wigner crystal at $\nu _{1}=0.2$ and of the
hole Wigner crystal at $\nu _{1}=3.8$ obtained by adding a filling fraction $%
0.2$ of holes to a fully filled (i.e., $\nu _{1}=4$) Landau level. All line
shapes are very distinct from that of the corresponding uniform states. The
crystals line shape tend to have long frequency tails and much larger width
than the uniform states (almost $\approx 14$ kHz in some cases). However,
this value is small in comparison with the width of the crystal line shape
in GaAs quantum wells\cite{Tiemann,CoteNMR} which is about $30$ kHz. The
hyperfine coupling is much larger in GaAs quantum wells than in graphene.

\begin{figure}[tbph]
\includegraphics[scale=0.8]{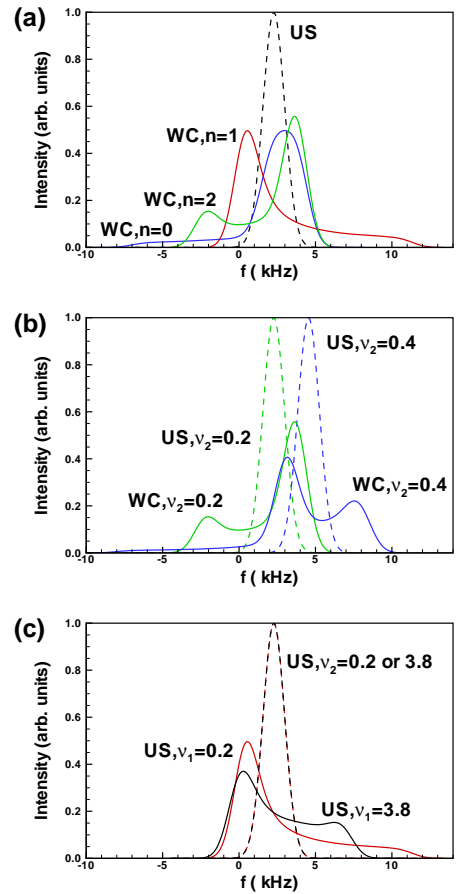}
\caption{(Color online)\ NMR\ line shapes of Wigner crystals (full lines)
and corresponding uniform states (dashed lines) at Zeeman coupling $\Delta
_{Z}=0.03$ for: (a) the electron Wigner crystal at filling factor $\protect%
\nu _{n}=0.2$ in Landau levels $n=0,1,2;$ (b) the Wigner crystal at $\protect%
\nu _{2}=0.2$ and the bubble crystal with 2 electrons at $\protect\nu %
_{2}=0.4$ in level $n=2;$ (c) the electron Wigner crystal at $\protect\nu %
_{2}=0.2$ and the hole Wigner crystal at $\protect\nu _{2}=3.8$, both for $%
n=2.$ The linewidth $\Gamma =1.0 $ kHz in all cases.}
\label{figure1}
\end{figure}

Contrary to the situation in GaAs quantum wells, there is no simple
relationship between the histogram of density (or, more precisely,
polarization) and the NMR\ line shape for the crystal states in graphene.
This is due to the fact that orbital and spin shifts are of the same order
and also to the fact that the orbital shift is not directly related to the
electronic density or polarization. The relative importance of the spin and
orbital shifts is shown in Fig. 2 where we plot the line shape obtained by
keeping only the spin shift, only the orbital shift, and both shifts in the
function $\Delta f\left( \mathbf{r}\right) $ in Eq. (\ref{rene7}) for the
Wigner crystal at $\nu _{1}=0.2.$ It follows that we cannot see sharp
differentiating features between, say, the electron and hole crystals shown
in Fig. 1(c). Moreover, we cannot give a simple explanation for the
difference between the Wigner and bubble crystals in Fig. 1(b).

\begin{figure}[tbph]
\includegraphics[scale=0.8]{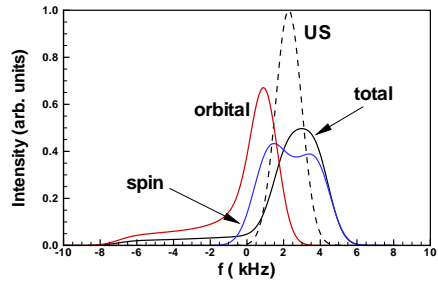}
\caption{(Color online) NMR line shape calculated with only the orbital
shift, only the spin shift, or with both shifts ("total" line) for the
Wigner crystal in Landau level $n=1$ at Zeeman coupling $\Delta _{Z}=0.03$
and with linewidth $\Gamma =1$ kHz.}
\label{figure2}
\end{figure}

For $\nu _{n}<1,$ only one valley (with spin down) is occupied by the
electrons$.$ In GaAs quantum wells, the ground state in $n=0$ around $\nu
_{0}=1$ is a crystal of spin skyrmions. In graphene, because of the valley
degeneracy and the finite value of the Zeeman coupling, valley pseudospin
skyrmions instead of spin skyrmions are expected to populate the ground
state for $\nu _{0}$ or $\nu _{1}$ near $1$. Actually, numerical
calculations show that merons (half a skyrmion\cite{Sondhi,Revueskyrmion})\
are preferred over skyrmions when a \textit{crystal} is formed so that for $%
\nu _{0}>0.5$, the electrons occupy both valleys leading to a state with
complete valley pseudospin depolarization. Such a meron crystal state is
depicted in Fig. 1 of Ref. \onlinecite{CoteSkyrmeGraphene} for the case $\nu
_{0}=0.8$ and reproduced in Fig. 4(b) below. The merons are arranged in a
checkerboard configuration with four merons per unit cell. Nearest neighbors
have opposite vorticity and opposite orientation of the valley pseudospin
component $P_{z}$ at their center. The electronic density is a
two-dimensional charge density wave$.$ The amplitude of the density
modulation is small. For example, for $\nu _{0}=0.6$, the density varies
between $0.5/2\pi \ell ^{2}$ and $0.7/2\pi \ell ^{2}.$ Since electrons
occupy the spin down levels only, the modulation of the spin component $%
S_{z} $ is also small. Consequently, the NMR\ line shape of the meron
crystal is not that different from that of the uniform state as can be seen
in Fig. 3(a). To see the detail of the modulations, a smaller value of $%
\Gamma $ is necessary. Fig. 3(b) shows the line shape from the same crystal
obtained with $\Gamma =0.2,\Gamma =1$ and $\Gamma =2$ kHz. The distinct
character of the meron crystal is visible for $\Gamma =0.2$ kHz while
crystal and uniform states have the same line shape for $\Gamma =2.$ In the
Wigner crystal (electron or hole) considered in Fig. 1, the spread in
frequency is larger and it is only for $\Gamma \approx 8$ kHz that the
difference between the two signals is lost. We remark that the same
checkerboard type of meron crystal is found at filling factor $\nu _{1}=0.8$
and the line shape is again close to that of the uniform state. In higher
Landau levels, this meron crystal disappears. For example, at $\nu _{2}=0.8,$
the ground state is a hole Wigner crystal. Our numerical approach, however,
does not allow us to compute the crystal state for $\left\vert \nu
_{n}-m\right\vert \lesssim 0.5$ for $m=1,2,3$ so that it is possible that
the meron crystal phase exists in a very narrow range around integer
filling. For isolated spin skyrmions, it was indeed shown in Ref. %
\onlinecite{Luo} that the filling factor range over which they exist in $%
\left\vert n\right\vert =1,2,3$ decreases rapidly with Landau level index.

\begin{figure}[tbph]
\includegraphics[scale=0.8]{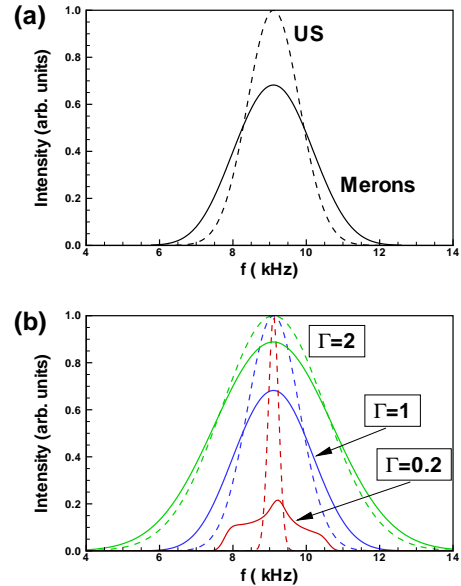}
\caption{(Color online)\ NMR\ line shapes of : (a) the meron crystal with
valley pseudospin texture in Landau level $n=1$ at filling factor $\protect%
\nu _{1}=0.8$ and Zeeman coupling $\Delta _{Z}=0.03$; (b) the crystal in (a)
calculated with different values $\Gamma $ of the linewidth as indicated.
The dashed lines give the line shapes for the corresponding uniform state in
each case.}
\label{figure3}
\end{figure}

A new crystalline ground state that we found when considering the spin
degree of freedom is a crystal with a skyrmionic spin texture. The
electronic density $n\left( \mathbf{r}\right) $ and the spin texture in the $%
x-y$ plane of this crystal are shown in Fig. 4 where $\Delta _{Z}=0.01,n=1$
and $\nu _{1}=2.2.$ In this state, level $\left( K_{+},-\right) $ is fully
filled while levels $\left( K_{-},-\right) $ and $\left( K_{-},+\right) $
are only partially filled, the former being more occupied than the latter
(an equivalent ground state is obtained by interchanging $K_{+}$ and $K_{-}$%
). The spin texture occurs in valley $K_{-}$ . In this state, there are two
vortices (the red square density patterns in Fig. 4) with the same vorticity
but differing by a global rotation of $\pi $ of the in-plane spin vector.
This configuration resembles the "square lattice antiferromagnetic (SLA)"
Skyrme crystal found in GaAs quantum wells around $\nu _{0}=1$\cite%
{BreySkyrme,Barrett} with the difference that the density profile on each
site now reflects the character of the two orbitals that enters the $n=1$
spinor in graphene which is given in Eq. (\ref{spinor}). A crystal with a
spin texture is possible only if the Zeeman coupling is small. Figure 5
shows the NMR\ line shape for the spin Skyrmion crystal in $n=1$ and at $\nu
_{1}=2.2$ at different values of the Zeeman coupling. For $\Delta _{Z}=0.03,$
[Fig. 5(a)], the spin texture is lost and the square lattice of Fig. 4 is
replaced by a triangular lattice of spin up electrons with filling factor $%
0.2$ on top of a uniform state (with filling $2.0$) of spin down electrons.
The corresponding NMR\ signal is similar to that of the Wigner crystals
shown in Fig. 1. At $\Delta _{Z}=0.02$ and $\Delta _{Z}=0.01,$ the ground
state is the spin skyrmion crystal of Fig. 4. The corresponding line shapes
are shown in Fig. 5. The line shape gets closer to that of the corresponding
uniform state (the dashed lines in Fig. 5) when the Zeeman coupling is
reduced. At $\Delta _{Z}=0.02,$ Fig. 5 shows the line shape of both the
triangular lattice (which is not the ground state) and the spin skyrmion
crystal ground state. They are almost identical.

\begin{figure}[tbph]
\includegraphics[scale=0.8]{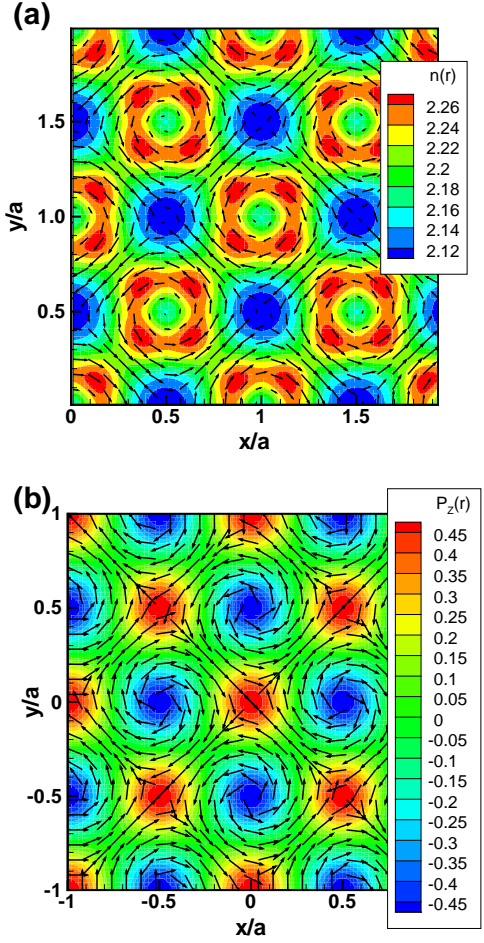}
\caption{(Color online) (a) Contour plot of the electronic density $n\left( 
\mathbf{r}\right) $ and spin texture in the spin skyrmion crystal in Landau
level $n=1$ at filling factor $\protect\nu _{1}=2.2$ for Zeeman coupling $%
\Delta _{Z}=0.01;$(b) contour plot of valley pseudospin component $%
P_{z}\left( \mathbf{r}\right) $ and valley pseudospin texture for the meron
crystal in Landau level $n=0$ at filling factor $\protect\nu _{0}=0.8.$}
\label{figure4}
\end{figure}

\begin{figure}[tbph]
\includegraphics[scale=0.8]{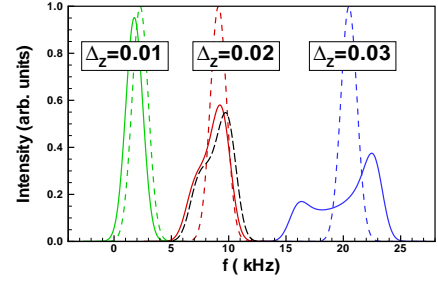}
\caption{(Color online) Evolution of the NMR\ line shape with Zeeman
coupling in Landau level $n=1$ at filling factor $\protect\nu _{1}=2.2:$
triangular Wigner crystal at $\Delta _{Z}=0.03;$ spin skyrmion and
triangular Wigner crystals at $\Delta _{Z}=0.02$ (full red line is the spin
skyrmion crystal and the dashed black line is the triangular Wigner
crystal); spin skyrmion crystal at $\Delta _{Z}=0.01.$ The linewidth $\Gamma
=1$ kHz in all cases. The dashed lines give the NMR signal from the
corresponding uniform states.}
\label{figure5}
\end{figure}

Figure 2 above shows that the spin and orbital shifts are equally important
in graphene according to our calculation which is based on a graphene
Hamiltonian written in the continuum approximation. This approximation,
however, does not capture the rapid oscillations of the electron wave
function near the carbon atoms. A first principle calculation would probably
be necessary in order to estimate the validity of our approach for the
orbital shift. We show in Fig. 6 the NMR\ line shape of the crystals
considered in Fig. 1(a) when only the spin shift is kept. The signal from
the crystal in $n=0$ is unchanged since there is no orbital shift in this
case. For $n=1,2$ the broadening of the line shape is reduced and the signal
is closer to that of the uniform state. Clearly, in graphene, the orbital
shift is important in order for the NMR technique to distinguish the crystal
from the uniform state.

\begin{figure}[tbph]
\includegraphics[scale=0.8]{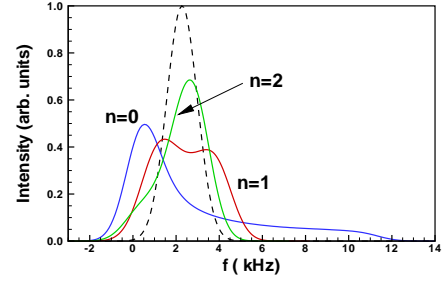}
\caption{(Color online)\ NMR\ line shapes calculated by keeping only the
spin shift of Wigner crystals (full lines) and corresponding uniform states
(dashed lines) at Zeeman coupling $\Delta _{Z}=0.03$ for the electron Wigner
crystal at filling factor $\protect\nu _{n}=0.2$ in Landau levels $n=0,1,2.$
The linewidth $\Gamma =1$ kHz.}
\label{figure6}
\end{figure}

The Hartree-Fock interaction in Eq. (\ref{fock}) is independent of the sign
of the Landau level index so that the phase diagram in levels $n<0$ is
identical to that in levels $n>0.$ Although the orbital shift in Eq. (\ref%
{rene15}) depends on the sign of the Landau level index, the line shape of
the crystals in levels $n>0$ can be simply related to the line shape of the
hole crystal in levels $n<0.$ For example, the electron Wigner crystal at $%
\nu _{1}=0.2$ has the same line shape as the hole Wigner crystal at $\nu
_{-1}=4-\nu _{1}=3.8.$ The spin polarization is the same in both crystals so
that the spin shift is the same. The orbital shift has also the same sign
since the current is carried by the holes for the crystal in $n=-1.$

We conclude this section by noticing that, for Landau level $n=1$ in
graphene, the relative nuclear frequency shift for an ion at $\mathbf{R}%
_{m}=0$ can be written as 
\begin{eqnarray}
\frac{\Delta f_{orb,1}}{\gamma _{n}/2\pi } &=&-1.\,\allowbreak 56\times
10^{-5}B_{0}\left[ \sum_{\xi ,\sigma }\sum_{\mathbf{p}}\right.  \label{orbi1}
\\
&&\times \left. \left\langle \rho _{\xi ,1,\sigma ;\xi ,1,\sigma }\left( 
\mathbf{p}\right) \right\rangle p\ell e^{-\frac{1}{4}p^{2}\ell ^{2}}\right] 
\notag
\end{eqnarray}%
using Eq. (\ref{rene15}). It is interesting to compare this result with that
for the 2DEG in GaAs. In this 2DEG, the orbital coupling hamiltonian is 
\begin{equation}
h_{orb}\left( \mathbf{r}\right) =\frac{\mu _{0}}{4\pi }\frac{e\hslash }{%
m^{\ast }}\gamma _{n}I_{z}\left( \frac{\mathbf{r}\times \mathbf{p}}{r^{3}}%
\right) _{z},
\end{equation}%
where $\mathbf{p}=m\mathbf{v}$ is the electron momentum and $m^{\ast
}=0.067m_{e}$ is the effective electronic mass. Following the same steps as
in Sec. 3(a) to compute the orbital shift, we find after a lengthy
calculation that 
\begin{eqnarray}
\frac{\Delta f_{orb,1}^{\text{GaAs}}}{\gamma _{n}^{\text{GaAs}}/2\pi }
&=&-4.\,\allowbreak 28\times 10^{-7}B_{0}^{3/2}\left[ \sum_{\sigma }\sum_{%
\mathbf{p}}\right.  \label{orbi2} \\
&&\times \left. \left\langle \rho _{1,\sigma ;1,\sigma }\left( \mathbf{p}%
\right) \right\rangle p\ell \left( 1-\frac{p^{2}\ell ^{2}}{\sqrt{2}+4}%
\right) e^{-\frac{p^{2}\ell ^{2}}{4}}\right] .  \notag
\end{eqnarray}
We checked numerically that the expression in the square brackets in Eqs. (%
\ref{orbi1}),(\ref{orbi2}) is of order $1$ for the crystal states so that
the orbital shift is at least an order of magnitude larger in graphene than
in GaAs at $B=10$ T. For $n=0,$ however, the relative frequency shift for
level $n=0$ is not zero in GaAs contrary to graphene. It is given by%
\begin{eqnarray}
\frac{\Delta f_{orb,0}^{\text{GaAs}}}{\gamma _{n}^{\text{GaAs}}/2\pi }
&=&-8.\,\allowbreak 20\times 10^{-7}B_{0}^{3/2}\left[ \sum_{\sigma }\sum_{%
\mathbf{p}}\right. \\
&&\times \left. p\ell \left\langle \rho _{0,0,\sigma }\left( \mathbf{p}%
\right) \right\rangle e^{-\frac{p^{2}\ell ^{2}}{4}}\right]  \notag
\end{eqnarray}%
which is however very small in comparison with the Knight shift in the GaAs
2DEG.

\section{CONCLUSION}

We have shown that in graphene both spin and orbital shifts contribute
equally to the NMR line shape of the crystal phases. From all the crystal
phases we studied, only the Wigner and bubble crystals have a line shape
that is distinctively different from that of the uniform state with the same
average filling factor. If our calculation does not overestimate the
contribution of the orbital shift, a measurement of the NMR\ line shape
would be one way to detect the formation of the electron crystal in
graphene. This conclusion, however, is valid as long as the linewidth of the
NMR\ signal in the uniform state is small. For the value of the hyperfine
spin coupling that we used in our numerical calculations, we estimate that
the maximal value of $\Gamma $ is around $8$ kHz. Since the spin and orbital
coupling parameters are not precisely known, this $\Gamma _{\max }$ must be
considered as only an estimate.

We have not considered the effect of quantum or thermal fluctuations in our
analysis. The importance of these fluctuations was discussed in detail in
connection with resistively-detected NMR\ measurement (RD-NMR) of Wigner
crystals in GaAs quantum wells\cite{Tiemann}. It is clear that they would
tend to wash out even more the difference between the line shape of the
crystal and that of the uniform state.

We assume in this paper that the NMR signal from the $^{13}$C
isotope-enriched graphene layer is sufficiently intense and/or the
experimental setup is sufficiently sensitive to allow the measurement of the
change in NMR\ line shape induced by the formation of the crystal states. As
discussed in Sec. V, this is a challenge for bulk NMR measurements where the
signal is proportional to the number of polarized nuclei which is very small
in a two-dimensional system such as graphene. A technique such as RD-NMR\
would probably be more appropriate for this type of measurement since it is
well suited for systems with a small number of nuclei\cite{Gervais}. Our
calculations of the NMR line shapes should apply to this technique as well.

\begin{acknowledgments}
R. C. was supported by a grant from the Natural Sciences and Engineering
Research Council of Canada (NSERC) and J.M. P. by scholarships from NSERC
and the Fonds de recherche nature et technologies du Qu\'{e}bec (FQRNT).
Computer time was provided by Calcul Qu\'{e}bec and Compute Canada.
\end{acknowledgments}

\end{document}